\documentclass[sort&compress,square,comma,authoryear]{aa}  

\usepackage{color}
\usepackage{graphicx}
\usepackage{graphicx}
\usepackage{dblfloatfix} 
\usepackage[symbol]{footmisc}
\usepackage{gensymb}
\def\SPSB#1#2{\rlap{\textsuperscript{\textcolor{black}{#1}}}\SB{#2}}

\def\SB#1{\textsubscript{\textcolor{black}{#1}}}
\usepackage{txfonts}
\begin{document}

   \title{A multi-scale exploration of a massive young stellar object -\\ a transition disk around G305.20+0.21?\footnotemark }
   \titlerunning{A multi-scale exploration of the MYSO G305.20+0.21}


   \author{A. J. Frost\inst{1}, R. D. Oudmaijer\inst{1}, W. J. de Wit \inst{2} \and S. L. Lumsden\inst{1}
          }

   \institute{School of Physics and Astronomy, University of Leeds, Leeds LS2 9JT, UK\\
              \email{pyajf@leeds.ac.uk}
         \and
             European Southern Observatory, Alonso de Cordova 3107, Vitacura, Santiago, Chile\\
             }

   \date{Received 6th November 2018 / Accepted 6th March 2019}

 
  \abstract
   {The rarity of young massive stars combined with the fact that they are often deeply embedded has limited the understanding of the formation of stars larger than 8M$_{\odot}$. Ground based mid-infrared (IR) interferometry is one way of securing the spatial resolution required to probe the circumstellar environments of massive young stellar objects (MYSOs). Given that the spatial-frequency coverage of such observations is often incomplete, direct-imaging can be supplementary to such a dataset. By consolidating these observations with modelling, the features of a massive protostellar environment can be constrained.}
   {This paper aims to detail the physical characteristics of the protostellar environment of the MYSO G305.20+0.21 at three size-scales by fitting one 2.5D radiative transfer model to three different types of observations simultaneously, providing an extensive view of the accreting regions of the MYSO.}
   {Interferometry, imaging and a multi-wavelength spectral energy distribution (SED) are combined to study G305.20+0.21. The high-resolution observations were obtained using the Very Large Telescope's MIDI and VISIR instruments, producing visibilities in the N-band and near-diffraction-limited imaging in the Q-band respectively. By fitting simulated observables, derived from the radiative transfer model, to our observations the properties of the MYSO are constrained.}
   {The VISIR image shows elongation at 100mas scales and also displays a degree of asymmetry. From the simulated observables derived from the radiative transfer model output we find that a central protostar with a luminosity of $\sim$5 $\times$ 10$^{4}$L$_{\odot}$ surrounded by a low-density bipolar cavity, a flared 1M$_{\odot}$ disk and an envelope is sufficient to fit all three types of observational data for G305.20+0.21. The weak silicate absorption feature within the SED requires low-density envelope cavities to be successfully fit and is an atypical characteristic in comparison to previously studied MYSOs.}
   {The fact that the presence of a dusty disk provides the best fit to the MIDI visibilities implies that this MYSO is following a scaled-up version of the low-mass star formation process. The low density, low extinction environment implies the object is a more evolved MYSO and this combined with large inner radius of the disk suggests that it could be an example of a transitional disk around an MYSO.}


\maketitle
%
\renewcommand{\thefootnote}{\fnsymbol{footnote}}
\footnotetext[1]{Based on observations made with ESO Telescopes at the Paranal Observatory under programme IDs 097.C-0320, 75.C-0755 and 74.C-0389}

\section{Introduction}
Massive stars ($\geq$8M$_{\odot}$) are important driving factors within our universe. On the local-scale, the winds, outflows and supernovae (SNe) of massive stars can both inhibit and trigger further stellar formation, replenishing and sustaining turbulence resulting in the compression and rarefaction of molecular clouds \citep{krum14}. Massive stars generate photons with energies high enough to ionise hydrogen atoms within their surrounding medium creating features unique to massive protostellar environments such as H{\sc ii} regions \citep{church}. On the larger scale, massive stars change interstellar chemistry, fusing helium into heavier elements through the slow and rapid nuclear processes. Their stellar winds and SNe then distribute this material throughout the interstellar medium, increasing its metallicity. Massive stars are additionally important for extragalactic studies, allowing the inference of the mass of spatially unresolved galaxies \citep{ken}. SNe ejecta and massive stellar winds are also very influential at these large scales and powerful enough to contribute to the galactic super-wind, with stars $\geq$50M$_\odot$ providing a critical role in the superwind topology \citep{leith}.

Despite their evident importance, the formation of massive stars is poorly understood as their embedded nature, distance and rarity have restricted observational studies. Low-mass young stellar objects (YSOs) evolve through the creation of disks, which appear as the cloud collapses to conserve angular momentum and eventually accrete material onto the star. A massive young stellar object (MYSO) is defined as an infrared-bright object that has a spectral energy distribution (SED) that peaks at approximately 100$\mu$m and a total luminosity of greater than 10$^{4}$L$_{\odot}$ and it was previously thought that the radiation pressure associated with these large luminosities could interfere with disk accretion. \citet{kraus} confirmed that an accretion disk could be present around a massive forming star, using K-band interferometric observations and radiative transfer modelling to detect a disk emitting in the near-infrared at scales of $\sim$10au around the MYSO IRAS 13481-6124. Disks are now being more commonly detected at larger scales, with \citet{john} and \citet{jilee16}, for example, presenting findings of Keplerian-type disks of 12M$_{\odot}$ and 1-2M$_{\odot}$ respectively around two MYSOs. The review by \citet{bw} presents a detailed examination of the types of the disk-like objects around varying sizes of protostar and their accretion properties and note that the nature of the inner regions of the disks of O and B-type stars are far less clear than those of A-type and lower-mass stars. Direct accretion events associated with massive stars have also been observed, with \citet{car18} inferring an accretion event of the order 10$^{-3}$M$_{\odot}$yr$^{-1}$ around the MYSO S255 NIRS 3 when a radio burst was observed after a maser/infrared accretion event.

By combining observations at multiple wavelengths, different scales of an astronomical object can be traced. SEDs are multi-wavelength observables, but only provide indirect spatial information. Spatially resolved observations are required to break the SED degeneracy and confirm the origin of protostellar emission. In the infrared (IR) wavelength regime one can obtain such observations; its short wavelength range benefits the angular resolution and traces material in protostellar environments at $\sim$100K. A number of previous works, such as \citet{wit07}, \citet{linz2008} and \citet{vehoff}, present studies of MYSOs with mid-infrared interferometry in particular. \citet{boley13} present a survey of 20 MYSOs with the MID-infrared Interferometer (MIDI) instrument, fitting geometric models to the central N-band frequency of 10.6$\mu$m. \citet{wit10} investigated the MYSO W33A by fitting a radiative transfer model to an SED and MIDI visibilities, concluding that the majority of the mid-infrared emission originates from the outflow cavity walls, in agreement with other works such as \citet{deb}. Studies have been done at longer infrared wavelengths with \citet{wheel} and \citet{witcomics} presenting surveys of MYSOs at $\sim$20 and 24.5$\mu$m respectively. In \citet{wit11}, a singular MIDI baseline and an image profile at 24.5$\mu$m are fit with a radiative transfer model to investigate the MYSO AFGL 2136 but given the limited amount of information retrieved from a single interferometric baseline, the constraint on the N-band geometry is therefore lacking. 

The work in this paper improves upon these aforementioned studies by simultaneously fitting multiple MIDI configurations, a near-diffraction-limited $\sim$20$\mu$m imaging profile and an SED. By finding a radiative transfer model that optimally fits the MIDI data at 10mas scales, the VISIR data at 100mas scales and the SED stretching from near-infrared to millimetre wavelengths, the characteristics and accreting regions of the protostellar environment are probed. 

\begin{table*}[h!]
\caption{MIDI observations of G305.20+0.21}              
\label{table:3}      
\centering                                      
\begin{tabular}{l l l l l l l}          
\hline\hline                        
Configuration & Date/time & Telescopes & Projected baseline &  Position angle & Avg. visibility & ESO Run ID  \\    
Label & (UTC) & & (m) & ($^{\circ{}}$) & & \\
\hline                                   
A & 2005-06-26 01:20 &  UT1-UT2  & 42.5  & 48.1  & 0.03  & 75.C-0755(B) \\
B & 2005-02-27 03:35  & UT2-UT3  & 44.7  & 4.8  & 0.03 &  74.C-0389(A) \\
C & 2005-03-02 05:26  & UT3-UT4 &  56.7  & 83.4 &  0.02  & 74.C-0389(B) \\
D & 2005-03-02 06:10  & UT3-UT4  & 58.9 &  93.4  & 0.03 &  74.C-0389(B) \\
E & 2005-03-02 07:13  & UT3-UT4  & 61.0  & 106.9 &  0.03 &  74.C-0389(B) \\
F & 2005-06-24 02:46 &  UT3-UT4 &  62.3  & 146.7  & 0.01 &  75.C-0755(A) \\
\hline                                             
\end{tabular}
\end{table*}

This paper presents our results when this method is applied to the MYSO G305.20+0.21. G305.20+0.21 (henceforth G305) is located at RA 13:11:10.45, DEC -62:34:38.6 (J2000 co-ordinates). The region that houses G305 is littered with Class II maser emission (e.g. \citet{maser2}, \citet{hind}), which has been shown to be exclusively associated with massive star-forming regions (e.g. \citet{vander}, \citet{pandian}). A separate compact H{\sc ii} region is present $\sim$18" west of the source \citep{aj}, again implying the presence of massive star formation. A very bright source is also visible $\sim$22" east of G305 in 870$\mu$m maps from the ATLASGAL survey \citep{sch}. \citet{ajw} detect $^{13}$CO and HCO+ line wings in the area surrounding G305 implying the presence of outflows, but due to the spatial resolution of their data they cannot discern which source would be powering these emission features. G305 is classified as an MYSO in the Red MSX Source (RMS) survey \citep{rmslum} and has a listed bolometric luminosity of 4.9$\times$10$^{4}$ L$_{\odot}$ \citep{rmsmott10} at a kinematic distance of 4kpc \citep{ur}. \citet{krish} determine a distance of 4.1\SPSB{+1.2}{-0.7} kpc for a maser parallax of 0.25$\pm$0.05mas. \citet{hind} find $v_{LSR}$ = -42.0 kms$^{-1}$ from observations of the NH$_{3}$(3, 3) line at 24 GHz, corresponding to a kinematic distance of 4.74$\pm$1.70 kpc. \citet{aj} find G305 to have a deeply embedded near-infrared source and determine its luminosity to be 2.5$\times$10$^{5}$L$_{\odot}$ at a distance of 4.5kpc based on IMF models and Class II maser emission. Despite this high luminosity they detect no radio emission from the source, conclude that the source is not yet producing ionising photons and that G305 is therefore an MYSO. \citet{mengyao} included G305 (named as G305B in their paper) in a SOFIA survey of massive star formation supplemented by radiative transfer modelling and we discuss our findings in relation to theirs in Section 5.1. The source was also included in a study of a number of suspected MYSOs by \citet{jilee13} where the CO band-head emission of the MYSO (as part of a large sample) was modelled and the presence of a disk inferred.

The remainder of this paper is split into five further sections; Section 2 - a detailing of the observations, Section 3 - a description of our modelling,  Section 4 - a presentation of the results for this object, Section 5 - subsequent discussions and Section 6 - our conclusions.


\section{Observations}

In this study of G305, two sets of spatially resolved data in the mid-infrared and an SED are presented. The MIDI data have previously been presented elsewhere and the VISIR data is newly obtained. We describe each of the data sets in the following subsections.

   \begin{figure*}[h!]
   \centering
   \includegraphics[width=160mm]{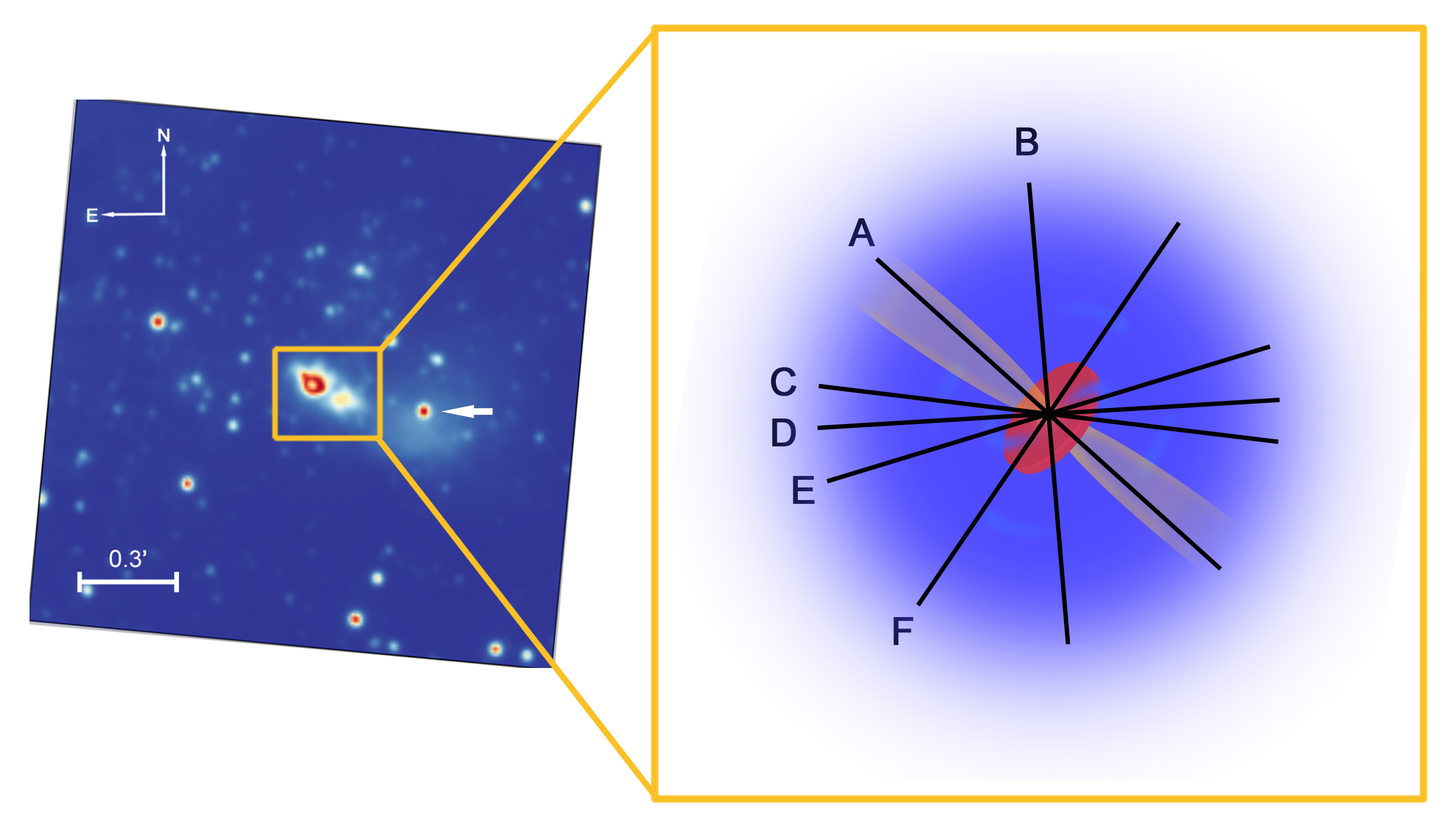}
   \caption{A logarithmically scaled VVV Ks-band image of G305.20+0.21 (left). Two lobes are visible separated by a dark lane. The bright point-source (noted with a white arrow) approximately 0.3' west of the central object surrounded is a known compact H{\sc ii} region. The Ks-band image is supplemented by a schematic of the suspected MYSO geometry overlain with the projected baselines of each configuration of the MIDI observations (right). In the schematic the envelope is shown in blue, the cavity in yellow and the disk in red.}
   \label{vvv}
   \end{figure*}

\subsection{Previously published interferometric data from MIDI}

MIDI \citep{midilol} is a Michelson-type, two-telescope interferometric instrument at the European Southern Observatory's (ESO) Very Large Telescope (VLT), able to combine light from the 8m Unit Telescopes (UTs) or the 1.8m Auxiliary Telescopes (ATs) into two interferometric channels. The channels have a phase difference of $\pi$ radians, induced by the half-reflecting combiner plate of the instrument, allowing the cancellation of uncorrelated background photons whilst retaining the correlated flux. The beam-combination is done near to the pupil plane while the signal is detected in an image plane. In order to minimise contamination of the data from background and instrumental thermal emission, the optics are cooled to 35K using a helium closed-cycle cryostat and the array detector is cooled to 10K. MIDI measures both the correlated and total flux in the N-band using one of two possible dispersive elements, grism or prism.

This work includes the reduced data from \citet{boley13} and samples of that data were re-reduced in spot checks to ensure consistency. The observations were taken in MIDI's HIGHSENS mode, where correlated and total flux measurements are taken separately and the photometric observations immediately followed the interferometric measurements to produce the final visibilities. HIGHSENS measurements are limited by the sky brightness variation between the photometric and interferometric measurements. The photometric measurements have up to a 15\% uncertainty that becomes a systematic uncertainty in the visibilities. Each target measurement was also preceded or followed by a calibration measurement of a star of known brightness and diameter. Further information on these observations is detailed at length in \citet{boley13} and further details of the MIDI observation process in general can be found in \citet{prz}, \citet{ches} and \citet{lein}. 

G305 was observed in 2005 in six distinct telescope configurations (the details of which are listed in Table 1) using three different telescope pairs. The prism mode was used for configurations A-D and F and the grism mode was used for configuration E. Each configuration has a different projected baseline and position angle. In order to help illustrate the regions of the protostellar environment each configuration may trace, Figure 1 is presented - a VVV (VISTA Variables in The Via Lactea, \citep{minni}) survey Ks-band image displaying G305.20+0.21 and its surrounding area, supplemented with a schematic labelled with each interferometric baseline. We include the Ks-band image for illustrative purposes only and do not include it in the fitting process as it presents another level of complication and the object's Ks-band flux is already considered in the SED. The VLTI's baseline capabilities range from approximately 40-120m and the observations for G305 were taken with baselines between $\sim$42-62m. This work attempts to model the entire N-band spectrum between 7 and 13 $\mu$m and as such the range in angular resolution varies from 12mas (for a 62m baseline at 7.5$\mu$m) to 31mas (for a 40m baseline at 12.5$\mu$m).

The MIDI measurements of G305 show that the object has a comparatively low correlated flux between 7 and 8$\mu$m in the context of the full sample of MYSOs from the \citet{boley13} sample. The average overall visibility is 0.03 and the visibility is not constant with wavelength but shows a clear chromatic dependency, in part because of the strongly varying opacity within the silicate absorption wavelength region at 9.7 micron.

MIDI was decommissioned in March 2015 to make way for the next generation of instruments on the VLTI.

\subsection{New 20$\mu$m imaging data from VISIR}

The VLT Imager and Spectrometer for the mid-InfraRed (VISIR, \citet{lag}) was used in service mode to observe G305 as a part of ESO run 097.C-0320(A) (PI: A. J. Frost). The instrument is mounted on the Cassegrain focus of UT3 of the VLT, providing near-diffraction-limited imaging at high sensitivity in three mid-infrared atmospheric windows. The observations presented here were taken with the Q-band filter which has a central wavelength of 19.5$\mu$m and a half-band-width of 0.4$\mu$m. Operating at this longer mid-infrared wavelength, VISIR traces cooler material within the protostellar environment compared to MIDI, particularly envelope emission as based on conclusions from \citet{wheel}. To reduce the amount of noise, the VISIR detector is cooled to $\sim$9K and observes with short exposure times. Given the large background fluctuations at mid-infrared wavelengths the chopping and nodding technique is used instead of flat-fielding, with chopping frequencies between 2-4Hz and amplitudes of $\sim$13" (see ESO's VISIR manual, \citet{lag}). G305.20+0.21 was observed on 6/7/2016 with good sky conditions (PWV was 0.78mm) at exposure times of 0.0114s with a total time-on-target of 45 minutes. The observations followed an upgrade for the instrument which improved its efficiency by a factor of six and increased its sensitivity \citep{visup}. The science verification observations post-upgrade were taken in February 2016, the same year as the presented observations. ESO pipelines (version 4.3.1) were used to reduce the data, accounting for chopping and nodding corrections and averaging the subsequent frames to form one image. HD 123139 was observed as PSF standard on the night of the observations with a measured FWHM of 0.48". The Graphical Astronomy and Image Analysis Tool (GAIA, \citet{gaia}) was used to perform aperture photometry on G305 and on three other calibrator objects (HD 169916, HD 163376 and HD 111915) observed at the VLT. The amount of counts detected within the apertures were compared to the recorded fluxes of the objects and the difference in aperture size was accounted for. This led to a resultant flux density for G305 of 138$\pm$7Jy at 19.5$\mu$m. 

\begin{table}
\caption{Fluxes used in the SED\textsuperscript{*}}              
\label{table:1}      
\centering                                      
\begin{tabular}{c c c}          
\hline\hline                        
Source & Wavelength ($\mu$m) & Flux (Jy)\\
\hline                                   
2MASS J-Band & 1.25 &  (3.21$\pm$0.09) $\times 10^{-3}$ \\
2MASS H-Band & 1.662  & (2.09$\pm$0.1)  $\times 10^{-2}$\\
2MASS Ks-Band & 2.159  & 0.113$\pm$0.004 \\
WISE &  3.4 & 0.293$\pm$0.005 \\
GLIMPSE & 3.6 & 2.23$\pm$0.07 \\
WISE & 4.6 & 5.71$\pm$0.2 \\
MSX & 14 & (1.10$\pm$0.07) $\times 10^{2}$  \\
OSCIR & 18 & 117.53$\pm$0.29 \\
VISIR & 19.5 & 138$\pm$7 \\
MSX & 21 & (2.66$\pm$0.2) $\times 10^{2}$ \\
\textit{IRAS} & 60 & (3.17$\pm$0.6) $\times 10^{3}$ \\
PACS & 70 & (1.14$\pm$0.003) $\times 10^{3}$ \\
\textit{IRAS} & 100 & (8.16$\pm$1) $\times 10^{3}$  \\
\textit{SIMBA} & 12000 & (18.5$\pm$0.04) \\
\hline                                             
\end{tabular}
\end{table}

\renewcommand{\thefootnote}{\fnsymbol{footnote}}
\footnotetext[1]{The italicised fluxes are those omitted from the fitting due to the considerations discussed in Section 2.3.}

\subsection{Spectral energy distribution (SED)}

G305's source fluxes were compiled from the Red MSX Source (RMS) Survey \citep{rmslum} and the literature to create an observational SED for the object. The included fluxes are 2MASS J, H and K band fluxes, the WISE 3.4 and 4.6$\mu$m fluxes, GLIMPSE data at 3.6$\mu$m, MSX 14 and 21$\mu$m points, the 18$\mu$m flux from \citet{debuizer} (taken with the OSCIR instrument at the Cerro Tololo Inter-American Observatory 4m Blanco Telescope) and the PACS 70$\mu$m flux (obtained through the \textit{Herschel}/PACS Point Source Catalog\textsuperscript{\dag}). No GLIMPSE 4.5$\mu$m flux is present in archival data for the source and the GLIMPSE 5.8$\mu$m image is saturated so neither are included. The WISE fluxes are the colour-corrected fluxes for a 200 K black-body emitter. Other wavelength data also exist in the form of the IRAS 60$\mu$m and 100$\mu$m data points and the SIMBA 1.2mm flux \citep{faundez2004}. We do not consider these in our fitting, however, due to the presence of the ATLASGAL core described in Section 1 whose emission is contaminating these measurements. As we cannot reliably separate the two sources, we discard all flux measurements obtained with apertures larger than 20". We include the 19.5$\mu$m VISIR flux mentioned in the previous section and the flux calibrated MIDI spectrum in the SED fitting. All the included SED fluxes are listed in Table 2.

\renewcommand{\thefootnote}{\fnsymbol{footnote}}
\footnotetext[2]{\url{http://archives.esac.esa.int/hsa/legacy/HPDP/PACS/PACS-P/PPSC/HPPSC_Explanatory_Supplement_v2.2.pdf}}

\section{Radiative Transfer Modelling}

The model used in this work follows the schematic of an MYSO defined in \citet{whit03}; a central protostar surrounded by an envelope, a bipolar outflow cavity and potentially a disk. This is a generalisation of the low-mass star formation geometry of Class I objects \citep{shu77}. Given the number of disks present in recent observations of MYSOs and the fact that outflows and dusty envelopes are nearly ubiquitous around MYSOs this is a sensible approximation of the protostellar environment.

\subsection{Description of the code}

The 2013 version of the radiative transfer code Hochunk, which is presented in \citet{whitney}, was used to perform the radiative transfer modelling in this study. Hochunk uses the radiative equilibrium routines developed by \citet{bjork} in a 3D spherical-polar grid code \citep{ww}. The code is of the Monte-Carlo style, where packets, or in this case "photons", from the centrally defined source propagate through a defined dusty structure in a random walk and the absorption, re-emission and scattering of those packets are calculated. The code produces results simultaneously at all inclinations and images can be produced at several wavelengths (which can be convolved with broadband filter functions, including 2MASS, \textit{Spitzer}, IRAC and \textit{Herschel}) for comparison to observations. In this work 4$\times$10$^{7}$ photons were used to generate images of sufficient quality for post-processing.

Two options exist for the modelling of the protostellar envelope; one described by a power-law and the other as a rotating and infalling sphere that defines the density of the envelope according to its infall rate (\citet{ulrich}, \citet{ter}). The outflow cavity can be defined in either a conical or polynomial shape, with the polynomial defined as:

   \begin{equation}
      z = a\varpi^{b}
   \end{equation}
   
where $\varpi$ = ($x^{2}$ + $y^{2}$)$^{\frac{1}{2}}$ and constitutes a region in the spherical envelope with different density properties. The disk description has been greatly expanded on from the previous version of the code but is still a flared dust disk overall. It is split into two parts, a small-grain disk and a large-grain disk, each of which have their own defined dust type, scale-height, inner radius and outer radius. One can decide in the overall disk description how much of the total disk mass lies within each of the disks. The disk can also harbour complex features such as warps, gaps and spirals, all of which can be individually defined, making using of the 3D capabilities of the code. These features are not included in this work and as such we refer to our model as 2.5D. Disk accretion can be included, with the released accretion luminosity adding to the central luminosity of the source. 

We used the dust-types suggested by \citet{whitney} for each part of our protostellar environment. The envelope dust grains are the same as those used in \citet{whit03}, with a ratio of total-to-selective extinction $r_{v}$ = 4.3 which is typical of star forming regions. The grains are also coated in water-ice mantles that make up 5\% of the grain radius. The large grains included in the disk correspond to Model 1 in \citet{wood02} and are made of amorphous carbon and silicates. Finally, the cavity dust is composed of dust in accordance with the average galactic ISM grain model by \citet{kim}. The 2013 version of the code allows for the inclusion of polycyclic aromatic hydrocarbons (PAHs) which are added to the aforementioned dust types and were not offered as an option in the previous version of the code. G305 does not display any PAH emission features in its spectrum and as such we removed this emission from the model.

\subsection{Simulating observations}

   \begin{figure*}[h!]
   \centering
   \includegraphics[width=184mm]{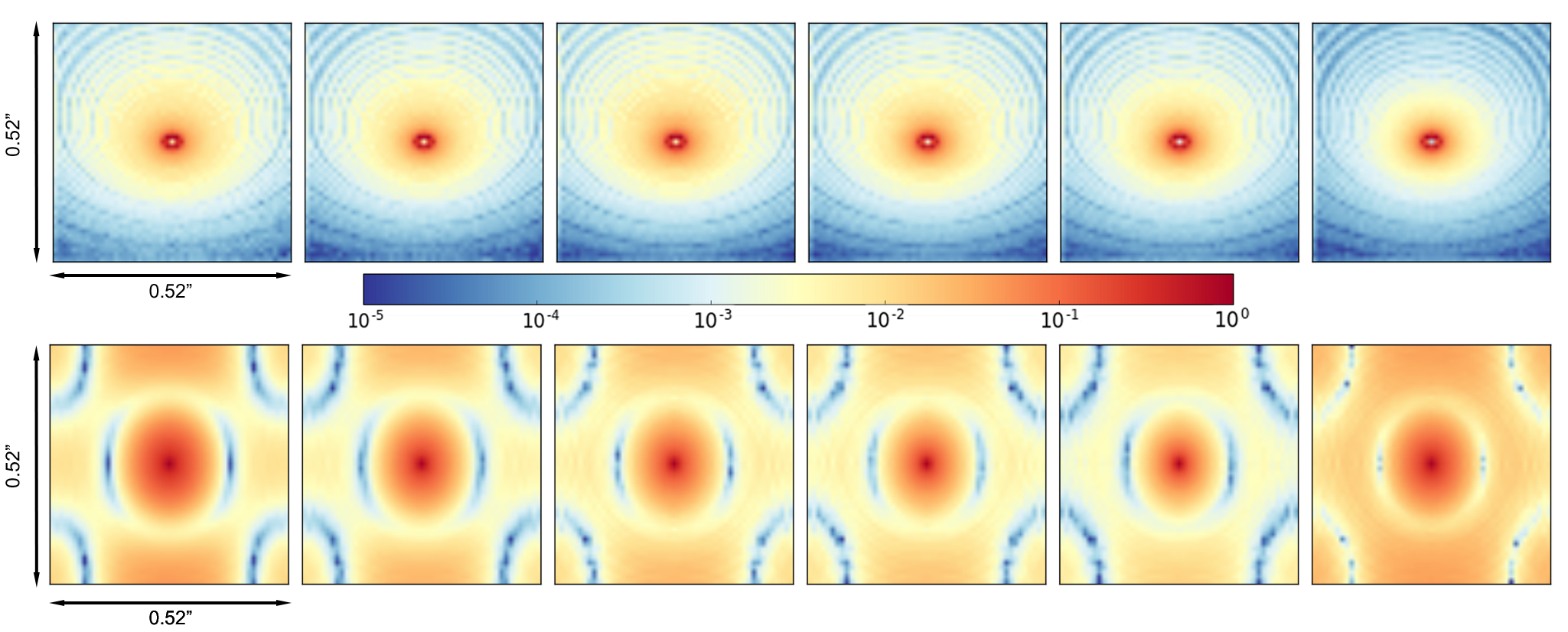}
   \caption{Logarithmically scaled model images (top row) at 7.5, 8.5, 9.5, 10.5, 11.5 and 12.5$\mu$m and their corresponding fast Fourier transforms (bottom row). Each image is 0.52" in width corresponding to the interferometric field of view. The colour bar applies to both sets of images and represents the number of counts.}
   \label{ffts}
   \end{figure*}

In order to simulate the MIDI observations, model images were generated at 7.5, 8.5, 9.5, 10.5, 11.5 and 12.5$\mu$m so as to sample the entire N-band wavelength regime. The images are 0.52" in size, corresponding to the MIDI slit width, with a pixel size of $\sim$0.01". In previous work (e.g. \citet{wit10}) the UT-airy disk was used as the field-of-view and the model images were multiplied with a Gaussian of identical size to this in order to account for the effects of the telescope. In this work we revised this methodology and have instead opted to use the slit-width as the size of our measurements and not to multiply by a Gaussian. This is because the MIDI data reduction process accounts for the effects of the telescope and removes them, meaning that multiplying the model images to include telescope effects is not necessary when the final aim is to simulate the visibilities, the final product of the data reduction. It is ultimately the slit size that determines the amount of observed emission from the object, and as such we use this as the size of our model images. We checked the result of multiplying the model image by a Gaussian corresponding to the UT-airy disk and note that it increases the resulting visibilities but not to such a degree that alternate conclusions about the MYSO geometry can be made.

Once the model images are generated they were post-processed in Python in order to allow a direct comparison to our observations. Taking the fast Fourier Transform (FFT) of the images provides a fully-filled u-v plane from which the visibilities corresponding to the position angle and baseline of the observations can be extracted. At lower densities gridding effects exist in the model images, creating ring-like structures within the cavity. In order to quantify the effect this may have on the final visibilities the model images were convolved with a Gaussian with a FWHM of 10 pixels to remove any high frequency structure and then subtracted from the original model image to isolate the rings and quantify their contribution to the emission. The rings constitute approximately 5\% of the overall image counts and therefore any effects they may have on the model visibilities are negligible. Example model images are presented in Figure 2, which also pertain to the best-fitting model discussed in Section 4.2.

In order to simulate the 19.5$\mu$m image a model image of the same field of view as our observations was generated ($\sim$14" across). This was convolved with the observed PSF standard star from the same night as the observations (HD 123139) in order to accurately include seeing effects on the modelled object. An azimuthally averaged profile of the model image was taken to allow easy comparison to the source and PSF.

To generate an SED the Hochunk post-processing script written by \citet{whitney} was utilised which extracts the flux and wavelength data from the radiative transfer output. Fluxes from the RMS and the MIDI observations were overlain on the model SED to allow comparison, with the beam-sizes and apertures taken into account in the SED generation. 
   
\subsection{Fitting process}

Given the very large parameter space of Hochunk it is not efficient to automatically fit all three data types and as such the following procedure was utilised. Since SEDs are degenerate (as previously discussed) and theoretically the easiest to fit, this was the starting point for the fitting process. Fitting the MIDI and VISIR data followed and the simulated visibilities and profiles were checked simultaneously as each model parameter was systematically varied. Extreme testing was done to determine whether a parameter was going to affect the datasets. If a change occurred in the simulated observables the values were changed in smaller intervals to then constrain the fit and improve the model. Given the success of the envelope-only model of \citet{wit10} in fitting the VISIR profiles of a range of MYSOs in \citet{wheel} and the MIDI visibilities of MYSO W33A \citep{wit10} the same central object (T = 35000K, R = 8R$_{\odot}$), envelope and cavity parameters were used as a starting point for the modelling of G305. However, the model did not immediately present a satisfactory fit and as such a large portion of the parameter space was varied in attempts to fit the observational data.

The VVV Ks-band image shown in Figure 1 displays a dark lane between two lobes. The lobes are likely to correspond to the bipolar cavities of the object, implying a dense dusty feature lies between them. This led to the inclusion of a disk in the model but we note that due to size of the dark lane ($\sim$2") it is more likely a result of the shadow of a disk on the surrounding nebulous material. Originally the presence of the lane also led to a prioritisation of close to edge-on inclinations for the source. However, close to edge-on inclinations do not reproduce well the 19.5$\mu$m image as too much emission is present in the model images from the red-shifted cavity lobe so ultimately less extreme inclination solutions were explored. The added disk was experimented with in a number of respects. Its mass was varied between 0.1-10 solar masses throughout the fitting. The scale-height of the disk and the scale-height exponent (which affect the curved shape of the disk at greater radii) were also experimented with. The outer radius of the disk was changed on 500-1000au scales in accordance with the expected radii of disks around other objects and the minimum disk radius was varied in intervals beginning with the sublimation radius (18au). The inclusion of a puffed-up inner rim and curved inner rim were also tested. The cavity properties such as shape, opening angle and density were varied as they were expected to affect the VISIR profiles. Varied envelope properties include the centrifugal radius, infall rate and its minimum and maximum radii. The inclination is a parameter that has the power to affect all the simulated observables and as such a compromise had to be made when deciding on the final value. A discussion of the effects of varying these parameters can be found in Section 4.1.

In an attempt to quantify the quality of our fits we calculated the chi-squared for each observation. The chi-squared was calculated separately for each configuration of the MIDI observations and each value combined to determine the best MIDI fit overall. The VISIR chi-squared was calculated from the radial profiles up to the cut-off displayed on the $x$-axis of the VISIR profiles (Figure 4, see Section 4), which is the point at which the MYSO profile falls to the level of the background noise. The SED chi-squared was calculated from six data points, corresponding to the filter wavelengths at which the post-processing script provides the user with the convolved model fluxes allowing for an accurate comparison to the observed fluxes. We do not, however, calculate an overall chi-squared for all three observations and decide on a best fit on the basis of overall minima. As the observations all trace different material, one can argue that finding an overall chi-squared for the three observations is not a particularly useful exercise. The chi-squared does present use in terms of narrowing down the closer fits, especially for the MIDI observations. A variety of different minimum radii for the disk and envelope presented similar fits to the MIDI data and the chi-squared value helped to constrain their values and determine the final best-fit. 

\section{Results}

\subsection{General observations}

This section discusses the observations made while experimenting with the parameter space of the radiative transfer code in the context of each probed scale of the MYSO. The distance is a parameter that affects all simulated observables. We assume the RMS distance of 4kpc for G305 and all the following observations are made for this specific distance.

\subsubsection{10mas scales}
The N-band emission depends on a wide number of model parameters, thus complicating the search for an optimal model. The inclusion of a circumstellar disk has an effect on the relative flux contributions of the geometrical structures and the visibilities, making them higher than an envelope-only model. This appears consistent as the disk constitutes a compact emitting region and adds a large amount of dense material close to the central object. As such one can expect the visibilities and fluxes to increase at the scales probed by our spatial frequencies, especially at the shorter MIDI wavelengths which trace hotter material. \citet{wit10} also find that adding a disk increases the visibilities at 7 and 8$\mu$m, and their simulated visibilities from Figure 10 of that work are very similar in shape to those presented here. 

Any changes to the disk of the object have repercussions on the simulated visibilities but do not affect the simulated VISIR profiles (see Section 4.1.2). The scale-height of the disk does not induce large effects and changing the scale-height exponent (which affects the curved shape of the disk at greater radii) also results in small effects. Varying the outer radius of the disk affects the simulated visibilities between 10-13$\mu$m only and effects are small between 1000, 2000 and 3000au. This outer radius affects the emission at longer millimetre wavelengths, which probe the cooler disk regions, but are not simulated in this work. The minimum disk radius has a significant effect on the simulated visibilities, producing a range of visibility values when the minimum radius is varied between $R_{sub}$-125au. This is due to the fact that a large amount of emission will be reprocessed by the inner rim presenting a turnover. As the minimum radius increases past 100au, the visibilities decrease as more dust is cleared from the inner protostellar regions and the model visibilities become comparable to a no disk-model.

Adding a puffed-up rim increases the visibilities slightly at all wavelengths for the smaller baseline configurations and at 7-8$\mu$m for all configurations. This appears consistent as the puffed-up rim provides a greater area of dust available for heating and subsequent emission at MIDI's observed wavelengths. Changing the inner rim shape from flat to curved results in marginally higher visibilities for Configuration A across the entire N-band spectrum. Since the disk is flared and the scale height at the inner rim is small, one would expect the change in visibilities to be small, if the disk was more toroidal then one would expect a bigger change as the curve of the inner rim would be more pronounced. 

Increasing the disk density by varying the disk mass from 1 to 5 and then 10 solar masses while maintaining other disk parameters was also investigated. This increases the visibilities at the extreme ends of the N-band range slightly with the most marked effects found at 7, 12 and 13$\mu$m, lowering the 7 and 8$\mu$m visibilities whilst raising the 13$\mu$m visibility, resulting in a flatter visibility spectrum. These wavelengths are the least affected by silicate absorption so one would expect a greater change in these regions of the spectrum. As the disk density increases the inner regions become optically thick extending the $\tau$=1 surface and lowering the visibilities. The improvements here come at the expense of the SED fit, which sees an increase in its 70$\mu$m peak and the fluxes between 1-7$\mu$m. The modelled N-band visibilities hardly change with the inclusion of disks lower than 1 solar mass. A 0.1 solar mass disk, i.e. a disk mass within the regime of those found in Herbig Be systems (e.g. \citet{herb1}, \citet{herb2}, \citet{herb3} and \citet{herb4}), was tried and this generated little change. From these observations we conclude that the N-band interferometry poorly constrains the total disk mass, which is in agreement with \citet{boley2016}, and conclude that the N-band visibilities are insensitive to most geometrical parameters of the disk for an object of this distance, except those related to the inner rim size and shape.

We note that throughout the fitting process Configuration A was consistently the most difficult to fit. In particular, the 7-8$\mu$m visibilities proved to be systematically inconsistent with the radiative transfer models that would match the other observables. This configuration has the smallest baseline and is therefore the lowest resolution configuration of the MIDI dataset. It is also the only one aligned with the outflow axis (as displayed in Figure 1). The model visibilities at 7-9$\mu$m are constantly higher than the observed visibilities meaning that the observed emission is less compact than the model predicts. Asymmetrical dusty structures along the outflow axis could arise from the entrainment of material during the accretion process. As dynamical processes that depend on local conditions are not covered by the radiative transfer model, we can only speculate about the emission's origin and more data are needed to interpret this plausibly. Another potential source of this emission is entrainment from a collimated jet. It is thought that collimated jets are associated with a number of MYSOs (e.g. \citet{si}, \citet{beut02}) and such a collimated jet would lie along the position angle of Configuration A. A jet is not included in our model and as such dusty material that might have been swept up into such a jet and emit would not be replicated in our simulated visibilities, again explaining the difficulties with fitting this configuration. Given these considerations, the fitting of the other configurations which are more likely to be accurately represented within the model were prioritised over fitting Configuration A.

\begin{figure*}[h!]
   \centering
   \includegraphics[width=180mm]{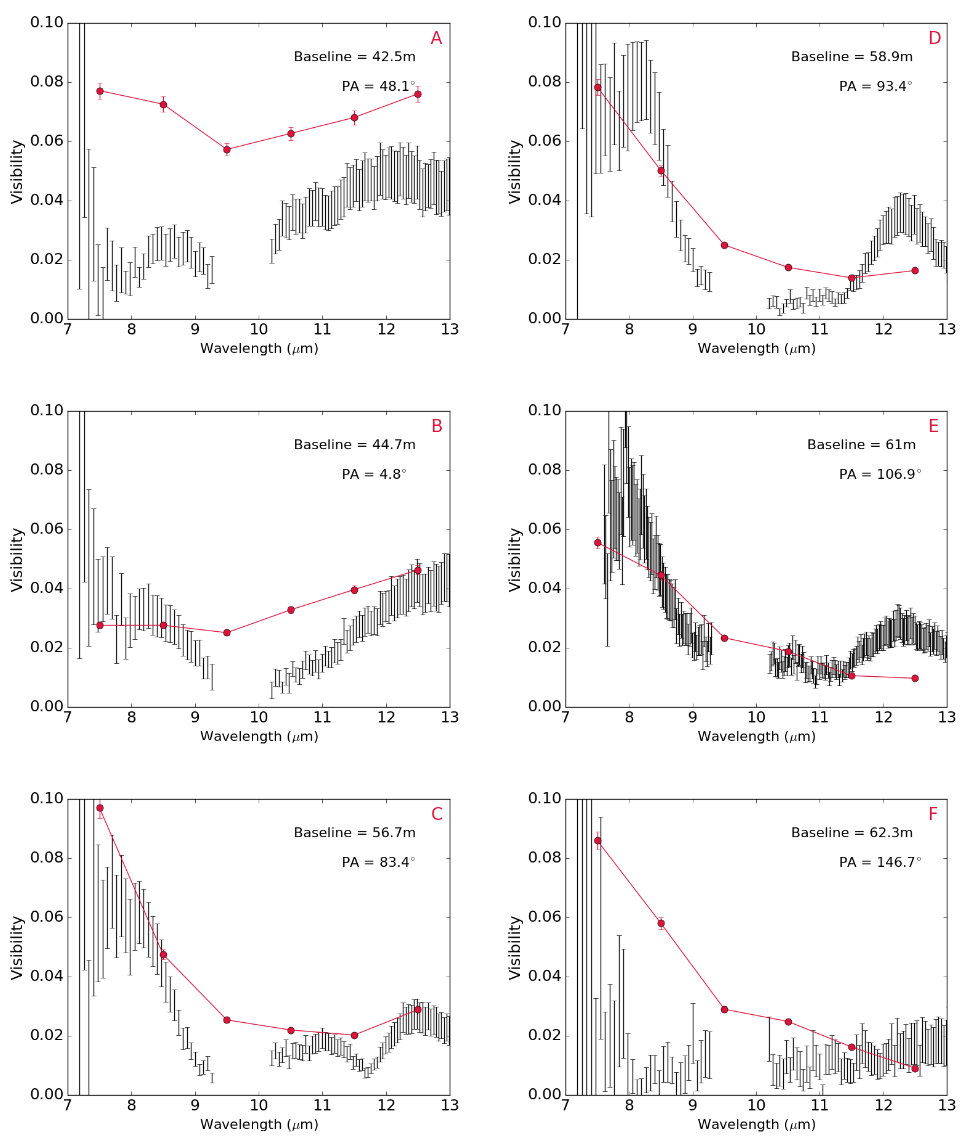}
   \caption{Observed visibilities for each configuration (black) with the simulated visibilities for each model image (red).}
   \label{vis}
   \end{figure*} 

\subsubsection{100mas scales}
All the disk parameters mentioned in the previous section have negligible effects on the simulated 19.5$\mu$m image. It was found that the most important influences on the emission at this wavelength are the cavity properties, in particular the cavity opening angle and its geometrical shape as captured by the cavity exponent ($b$ in Equation 1). Since the cavity density is low, the material within it will be more easily illuminated by the central object and the envelope material in the cavity walls will bear the brunt of the stellar emission and be heated. Therefore, the more emission from the cavity that is present, the more extended the emission in the Q-band becomes and the greater the amount of $\sim$20$\mu$m flux. When the cavity exponent $b$ equals 1 it will have a conical shape as opposed the more curved shape of the polynomial cavity that has an exponent other than 1. Because of this change in shape there is less flux at larger scales when modelling a conical cavity as it is essentially narrower, presenting a smaller area of cavity wall available for irradiation and subsequently a smaller solid angle on the sky for re-emission. The envelope infall rate also affects the $\sim$20$\mu$m emission. Increasing the envelope infall rate reduces the flux at larger radii. This appears consistent as more of the envelope mass will be present at smaller radii and within a region that is traced at mid-infrared wavelengths, as opposed to remaining in cooler outer regions that would be better traced by radio and sub-mm data. The inclination is the remaining parameter to present a significant effect on the $\sim$20$\mu$m emission but also affects all types of simulated observation. The affect of the envelope infall rate is less pronounced than the affects of varying the nature of the cavity walls and as such we conclude that VISIR is predominantly a tracer of cavity wall emission (within the limitations of our adopted dust radiative transfer model). 

The remaining factor that influences the $\sim$20$\mu$m emission is the inclination, but this affects all types of observation. Pushing to pole-on inclinations raises the MIDI visibilities, while decreasing the amount of $\sim$20$\mu$m flux (as one sees less of the illuminated cavity wall) and weakening the silicate absorption feature (as the cavity is optically thin) in the SED and as such a compromise needed to be found throughout the fitting. If the inclination is increased towards 50$^{\circ}$ for this source, more 19.5$\mu$m emission from the red-shifted cavity lobe is visible in the model images that is not observed in the source. The ultimately adopted value for the inclination of the modelled object is 35$^{\circ}$. This lies within the range of inclinations found from the five best fitting models of \citet{mengyao} and is similar to the inclination derived in \citet{jilee13} of 43$^{\circ}$. 35$^{\circ}$ was chosen as the final value because for higher inclinations the flux short-ward of the silicate absorption feature was too great and the depth of the silicate absorption feature in the SED was slightly too deep.

      \begin{figure*}[h!]
   \centering
   \includegraphics[width=150mm]{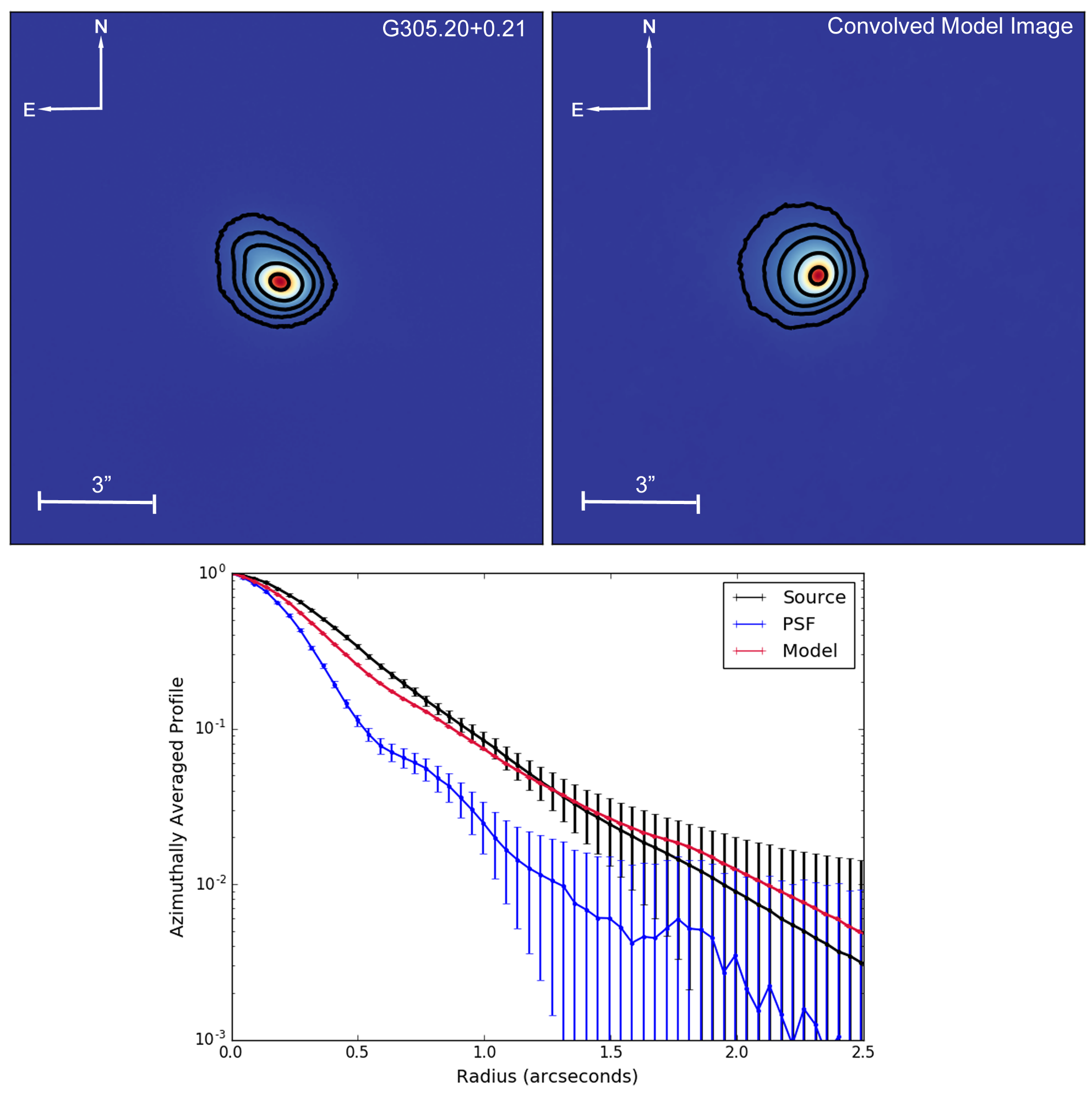}
   \caption{VISIR 19.5$\mu$m image (top left), convolved model image (top right) and subsequent radial profiles (bottom). The model image was convolved with the PSF of the observed object to accurately mimic the effects of the telescope specific to the observations.}
   \label{imgs}
   \end{figure*} 
   
         \begin{figure}[h!]
   \centering
   \includegraphics[width=90mm]{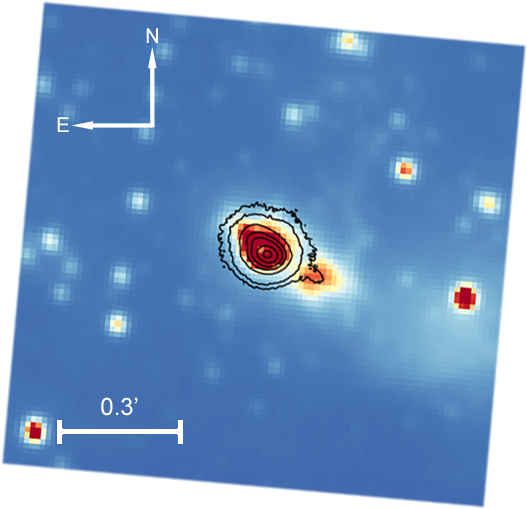}
   \caption{A combined image showing the location and morphology of our VISIR 19.5$\mu$m image (black contours) in comparison to the structure observed in the VVV K-s band image. The VISIR contours are 0.2, 0.5, 2, 5, 10, 25 and 75\% of the peak flux.}
   \label{overlay}
   \end{figure}

\subsubsection{The total flux}

Factors that can change the shape of the whole model SED are the size of the central illuminating star (mass, radius and temperature are the definable parameters within the code), the inclination and the distance. The inclination is also one of many parameters that affects the silicate absorption feature at 9.7$\mu$m. As the inclination is increased from a pole-on inclination (0 degrees) to an edge-on inclination (90 degrees) the silicate absorption feature deepens (an illustration of which can be found in \citet{whit03}). This is because edge-on inclinations probe the densest regions of the infalling envelope and disk of the MYSO, while pole-on inclinations probe the much less dense outflow cavity. MYSOs are known to be very dusty objects so one would expect the silicate absorption feature to be very prominent, but the N-band spectrum for G305 displays a weak silicate absorption feature. This implies that a closer to pole-on inclination for the model would provide a satisfactory fit to this feature. However, the Ks-band VVV image implies the presence of a second lobe and as such higher inclinations were trialled leading to a final inclination of 35$^{\circ}$.

The silicate absorption feature has long been an indicator of the extinction towards a source (e.g. \citet{henning}, \citet{boley2016}) and the interstellar extinction can be varied within the SED post-processing. A small amount of interstellar extinction best satisfies the model SED's silicate absorption feature, allowing it to remain small enough to fit the observed MIDI total fluxes. The appearance of two cavity lobes in the Ks-band already implies that the total extinction towards the source is lower than the average found in MYSOs, and when compared to other MYSOs such as AFGL 2136 \citep{wit11} the silicate absorption feature is indeed much weaker. This implies that most of the compact (<1") N-band emission suffers very mildly from foreground dust extinction, explaining why both outflow cones are detected in scattered light on scales of several arcseconds. \citet{boley13} fit a curve to the N-band spectrum, and find an optical depth of $\tau_{10.2}$ = 0.5 in the total flux and a depth of $\tau_{10.2}$ = 1.71 in the correlated flux, averaged over all measurements. They note that the optical depth of the silicate absorption feature can be significantly lower in the total flux than the correlated flux and postulate that this could be due to the fact that the total flux contains contributions of emission from all scales, while the spatial filtering of MIDI limits this contribution. They fit a power-law spectrum with an additional foreground screen of material with a column density and dust composition (independent of projected baseline or position angle) to the correlated flux for each of their sources. They found that G305 was one of the few sources whose correlated flux was not adequately reproduced by this model implying that the absorbing material is not largely uniform, detached from the scales traced by MIDI and that spatial effects may be present.

\begin{table*}[h!]
\caption{Parameters of the preferred model\textsuperscript{\ddag}}        
\label{table:2}      
\centering                                      
\begin{tabular}{c c c c c c c c c c c c c c c}          
\hline\hline                        
$M_\star$ & $L_\star$ & Temp. & $i$ & $d$ & $R\SPSB{min}{env}$ & $R\SPSB{max}{env}$ & $R_{c}$ & $\dot{M}_{infall}$ & $\theta_{cav}$ & $n_{cav}$ & $M_{disk}$ & $R\SPSB{min}{disk}$ & $R\SPSB{max}{disk}$ & $A\SPSB{for}{v}$ \\    
($M_{\odot}$) & ($L_{\odot}$) & (K) & ($^{\circ{}}$) & (kpc) & (au) & (au) & (au) & ($M_{\odot}yr^{-1}$) & ($^{\circ{}}$) & (gcm$^{-3}$) & ($M_{\odot}$) & (au) & (au) &  \\ 
\hline                                   
25 & 48500 & 35000 & 35 & 4 & 60 & 5$\times$ 10$^{5}$ & 2000 & 7.5 $\times$ $10^{-4}$ & 12 & 8.35 $\times$ 10$^{-21}$ & 1 & 60 & 2000 & 1 \\
\hline                                             
\end{tabular}
\end{table*} 

The total cavity optical depth in the line of sight (or the cavity density and path length) also affects the silicate absorption feature. The cavity density used in \citet{wit10} of approximately 1$\times$10$^{-18}$gcm$^{-3}$ results in a very deep silicate absorption feature (as observed in the case of the MYSO W33A) and our final model uses a lower density. Making the cavity opening angle greater also increases the depth of the silicate absorption feature in the model SED and as such the largest the cavity opening half-angle could be was 12$^{\circ}$. This is different to the half-angles found by the fitting \citet{mengyao} of $\geq$27$^{\circ}$ and we discuss this in Section 5.11.

The envelope parameters affect the SED across its wavelength range. Increasing the infall rate of the envelope deepens the silicate absorption feature and also increases the 70$\mu$m peak of the model SED. Increasing the maximum radius of the envelope factor shifts the entire SED to longer wavelengths and severely lowers the 70$\mu$m peak of the SED when pushed to sizes of order 10$^{7}$au (a value tried during the extreme testing mentioned in Section 3.3). The 70$\mu$m peak is an important photometric point to fit as it essentially describes the total flux of the MYSO. The 70$\mu$m flux increases because the optical thickness increases, and the colder part of the emission becomes more substantial. Making the disk around the central object more massive slightly increases the flux in the J, H and K bands and results in increases of the flux at the SED peak. As such a lower mass disk provided a more satisfactory fit.

\subsection{Best-fitting model} 

The systematic variation of the properties discussed in the previous section and the consideration of the chi-squared values for the MIDI fits finally amounts to the best-fitting model whose parameters are shown in Table 3. 
The central protostellar object of this model possesses a luminosity of 4.85$\times$10$^{4}$L$_{\odot}$ for the RMS distance of 4kpc, which is in very good agreement with the RMS bolometric luminosity of 4.9$\times$10$^{4}$L$_{\odot}$. In order to satisfy the MIDI visibilities, a disk of order $\sim$1 solar mass (the exact mass is poorly constrained as discussed in Section 4.1.1) is required ranging in size from 60-2000au in radius. The envelope has a centrifugal radius of 2000au, chosen to be consistent with the outer radius of the disk, and the infall rate of the envelope is 7.4$\times$10$^{-4}$M$_{\odot}$yr$^{-1}$. The cavity opening angle is 12$^{\circ}$ and the cavity density is 8.35$\times$10$^{-21}$gcm$^{-3}$.

Figure 3 presents the MIDI visibilities for each observed configuration of G305, with the observed visibilities being shown in black and the simulated model visibilities being shown in red. For all baselines, we see a depression in the visibility spectrum around the silicate absorption indicating that the emission region at around those wavelengths is larger than the adjacent N-band wavelengths. Telluric absorption impairs flux measurements between 9.3$\mu$m and 10.2$\mu$m and as such they have been omitted. Any observations below 7.5$\mu$m or above 13.3$\mu$m are outside of the N-band and compromised by atmospheric absorption, so our visibilities are plotted between these values only. The errorbars of the simulated visibility points account for the error induced by the artificial ring structures discussed in Section 3.3. The emission seen in Figure 2 is relatively extended resulting in the low visibilities seen in Figure 3. The images also suggest that the inner regions of the cavity walls become progressively fainter with wavelength (with the exception of the 9.5 and 10.5$\mu$m images) with an increase in the typical emitting size and the subsequent production of lower visibilities.

Figure 4 shows the VISIR image for G305, the convolved model image and their subsequent radial profiles. The contours represent 2, 5, 10, 25 and 75\% of the peak flux and the error bars represent the rms within a given annulus. The secondary lobe seen in the VVV Ks-band image is also visible in the VISIR image (albeit it at low percentage contours of $\sim$0.2\% of the peak flux). Figure 5 presents a combination of the two images, illustrating that their emission coincides and confirming the location of the secondary lobe. When a Ks-band model image was generated for the source the secondary lobe was not visible in the normal intensity images. However, images detailing the polarised emission, images of the polarised flux and the Q and U Stokes images can also be generated. In these images the second lobe is visible, implying that most of the Ks-band emission from the secondary lobe in the VVV image is scattered light. 

Figure 6 shows the SED. The fluxes corresponding to the MIDI fluxes are shown as red crosses, included to assure consideration of the silicate absorption feature at 9.7$\mu$m. The blue diamonds are fluxes from the RMS, the yellow diamond represents the VISIR flux and unfilled green diamonds represent the fluxes from the RMS survey which were omitted due to the considerations made in Section 2.3. We note that the near-infrared fluxes show a slightly worse fit than the rest of the SED. G305 has an envelope which will be causing significant levels of extinction and in reality this envelope is unlikely to be as smooth as the one presented by the model. The envelope extinction effects will have a much larger effect on the near-infrared fluxes than others at longer wavelengths in the SED. Given this and the potential for irregularities within the envelope material, fitting of the rest of the SED was prioritised over fitting the near-infrared fluxes.

\renewcommand{\thefootnote}{\fnsymbol{footnote}}
\footnotetext[3]{The level of constraint on each of the quoted values is discussed in Section 4.}

      \begin{figure}[h!]
   \centering
   \includegraphics[width=90mm]{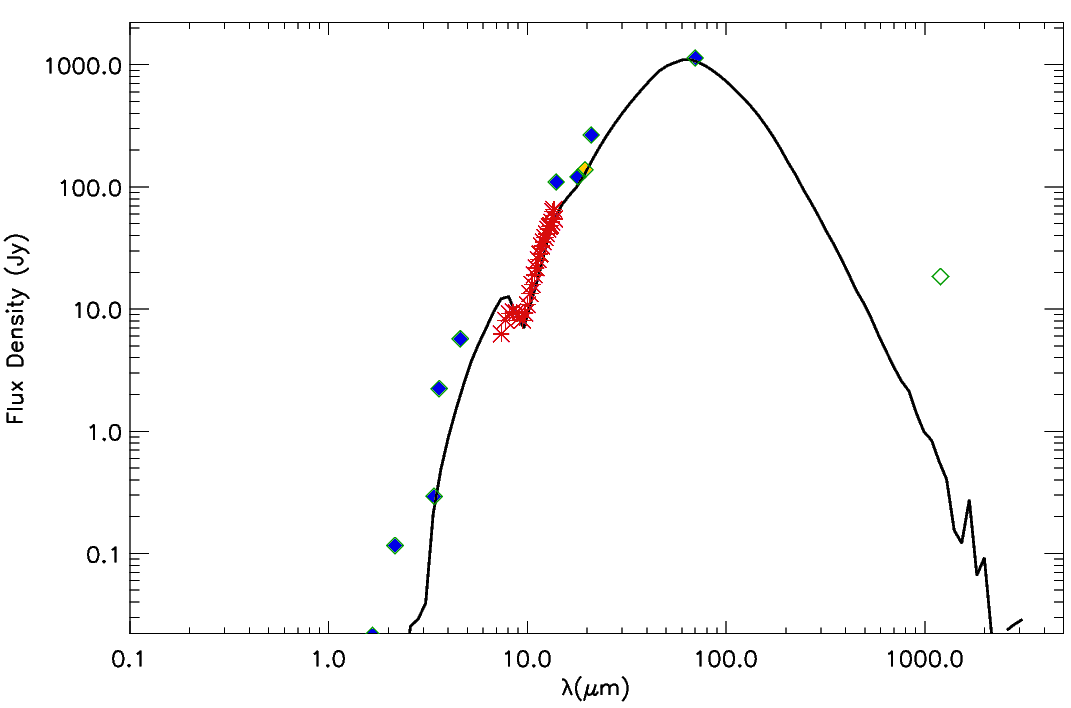}
   \caption{Model spectral energy distribution of the best-fitting model (black). Multi-wavelength flux measurements from the RMS are represented as blue diamonds, the yellow diamond represents the VISIR flux density and the fluxes corresponding to the MIDI visibilities are also shown in red. The unfilled diamonds represent the fluxes that were not considered in the fitting due to their suspected contamination.}
   \label{sed}
   \end{figure}

A temperature/density map of the modelled final MYSO is presented in Figure 7, displaying the simulated object at scales corresponding to the three different kinds of observation. The final schematic for the object consists of a dusty envelope with a low-density bipolar cavity carved out by the central protostar and a large-grain disk. We do not include a small grains disk in the model as it presents another layer of complexity which is not conducive to constraining a good fit. The parameters of the disk were decided in such a way as to match a density feature induced by the Ulrich-type envelope. As the envelope rotates, material will collect at the centrifugal radius as it falls towards the central object, leading to a bulge that is visible in the central density map of Figure 7. Since such a feature would lead to the creation of the disk, we fixed the scale-height of the disk to match this density feature to ensure consistency.  

   \begin{figure*}[h]
   \centering
   \includegraphics[width=180mm]{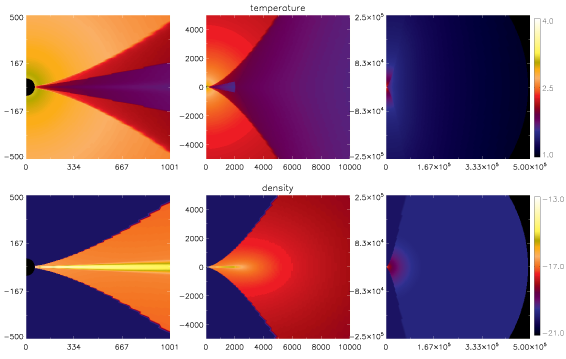}
   \caption{Cut-through, side-on, logarithmically-scaled maps of the temperature and density for the best-fitting model generated with Hochunk (plotted using IDL\textsuperscript{\S} post-processing scripts written by \citet{whitney}). The temperature colour-bar is in Kelvin and the density colour-bar is in gcm$^{-3}$. The axes for all panels are in au. Each column of images represents a different scale, corresponding to the observations, with the smallest scales traced by MIDI on the left, larger scales traced by VISIR in the centre and the entire envelope displayed on the right.}
   \label{tempdens}
   \end{figure*}
   

\section{Discussion}

\subsection{Comparisons with previous work}

The most notable difference between this final model and the starting point, the envelope-only model of \citet{wit10}, is the addition of a disk. Other significant departures from the envelope specification are the centrifugal radius, which is 2000au as opposed to 33au, changed to match the outer radius of the added disk. The cavity has the same polynomial shape as de Wit's work, but in order to successfully fit the silicate absorption feature of the SED the cavity density in our best-fitting model is lower than that presented in \citet{wit10}. The central protostar is 30\% dimmer than the central protostar required to fit W33A. The fact that W33A does not require a disk to fit its MIDI visibilities while G305 does could be due to their environments. \citet{wit10} note that while a disk was not visible for W33A, MYSOs viewed at smaller inclinations may reveal the presence of a dusty accretion disk and this could be the case for G305. More recently, \citet{maud} show that at larger scales W33As protostellar environment is very chaotic, with the spiral-like structures wreathing their way around the central source suggesting a highly turbulent, disturbed environment. This is the antithesis of the environment of G305, which appears to be a very `typical' YSO with clearly defined outflow cavities and a disk visible even in near-infrared images. One explanation for the difference between the two MYSOs could be that W33A could have formed a disk like G305 that was then was either disturbed post-formation by the processes creating the spiral structures mentioned above, or alternatively that the system has always been so volatile that the disk was never able to form at these scales.

\citet{wheel} compare $\sim$20$\mu$m profiles generated from the envelope model of \citet{wit10} to the VISIR images of a number of MYSOs. They deduce that this model can reproduce the images and SEDs for the majority of their sample. Adding MIDI, as done in this work, adds a stringent constraint to the interpretation of the geometry of the source. Our analysis highlights the fact that the VISIR data is not sensitive to the disk properties of the YSO but is heavily influenced by the cavity properties, while the MIDI is affected by both. Given this, it is therefore unsurprising that the envelope only model fits the majority of their sample and this does not rule out the presence of disks around the MYSOs. MIDI probes the MYSO at scales small enough to trace disk emission, while the sensitivity of VISIR is not high enough for such an endeavour given the MYSOs distances. This indicates that if one wants want to identify the presence of a disk, whilst also accurately constraining the cavity properties of MYSOs in the infrared, a combination of the two observations is invaluable.

As mentioned briefly in the introduction, G305 has also been included in a survey of massive star formation by \citet{mengyao}. In this work an SED of G305 was compiled from fluxes obtained from Gemini/T-ReCS and SOFIA/FORCAST images and was fit using their ZT radiative transfer models \citep{zt} resulting in five best-fitting models. The stellar mass and inclination of our preferred fit are comparable to the ranges for these values listed for their five best fitting models. Other notable parameters such as cavity opening angle differ between our two works, but we note that the spatial resolution of our N-band data is higher and that therefore we were more tightly constrained on this parameter.

Within their deep, single-dish, 3.8-24.5$\mu$m images a secondary source (referred to as G305B2) can be seen $\sim$1" to the north-east of G305 whose nature remains unclear. Interestingly, G305B2 is not present across the entire wavelength range, being absent from the shorter wavelength images, with an increasing relative brightness throughout the N-band and progressively becoming less visible again towards 24.5$\mu$m. \citet{mengyao} mark a silhouette in their 9.7$\mu$m image between G305B2 and G305 and postulate that it could be indicative of the silhouette of an inclined disk. If this is a disk silhouette, that would imply that the inclination of the source is the opposite to what we find in this work. Changing the inclination of the model to its opposite (145$^{\circ}$ instead of 35$^{\circ}$) has minimal effects on the MIDI visibilities (as FFTs are symmetric), the VISIR radial profile (which is azimuthally averaged) and the SED. However it does present differences for our images, as the position of the brightest lobe is consistent only if the inclination is 35$^{\circ}$ for both the VISIR image and the VVV image (illustrated in Figure 5). The cavity lobe to the south-west is dimmer, and for a reverse inclination this would be the brighter lobe. This cavity lobe is outside the Gemini imaging field-of-view shown in \citet{mengyao} and consequently was not discussed in their work. We also note that asymmetry is present in our 19.5$\mu$m VISIR image and the Ks-band image and as such it is possible that they trace this second source.

In Section 4.1.1 we discussed the poor fitting of our MIDI data for Configuration A. The model visibilities were much higher than those observed, implying that extra emission must be present that is not being produced in the model. Given that the secondary source found in the Gemini images lies in the region of the suspected outflow cavity, it therefore appears a good candidate for the source of this emission. Whilst our MIDI data trace only a 0.52" area, one can assume that if material has been entrained out to nearly 1.5", material will be also be present at $\sim$0.5". 

Given the confirmed inclination of the source from the VVV and VISIR images, we conclude that G305B2 is unlikely to be a coherent disk structure. The point-like material could be entrained cavity material (as discussed in Section 4.1.1), fall-out from potential disk fragmentation or indicate the presence of a binary companion. The latter two possibilities are addressed further in Section 5.5.

\subsection{Definitely a disk?}

In Section 3.3 we mentioned that the unsatisfactory fits presented by the envelope-only model from \citet{wit10} and the fact that a dark lane existed between two apparent cavity lobes in the VVV Ks-band image led us to include a disk in the model. However, other studies find an insensitivity to disk emission and have shown that cavity emission alone can control the N-band visibilities (e.g. \citet{wit10}). As such, for completeness, Figure 8 is presented, which includes the best fitting model visibilities and those for the same model with no disk. The VISIR and SED fits for the no-disk model are not added as these change minimally with the inclusion or exclusion of a disk (see Sections 4.1.2 and 4.1.3). When the disk is removed the shape of almost all the model visibilities violates what is observed, especially the peak-like features observed in Configurations C, D and E between 7-8$\mu$m. This implies that the illuminated cavity walls represent too large a surface and that a compact component is required to recreate these visibilities. Configuration F is the only configuration satisfied by the no-disk model and constitutes the worst quality dataset of the group. We have already stated that to satisfy the silicate absorption feature within the SED that the cavity opening angle could not be larger than 12 degrees. Varying the cavity opening angle is the most effective way of changing the model visibilities in an envelope-only model, and with this limitation in place from the SED an envelope-only model could therefore not be found which would satisfy the N-band data as well as the present disk-including model.

\begin{figure*}[h!]
\centering
\includegraphics[width=180mm]{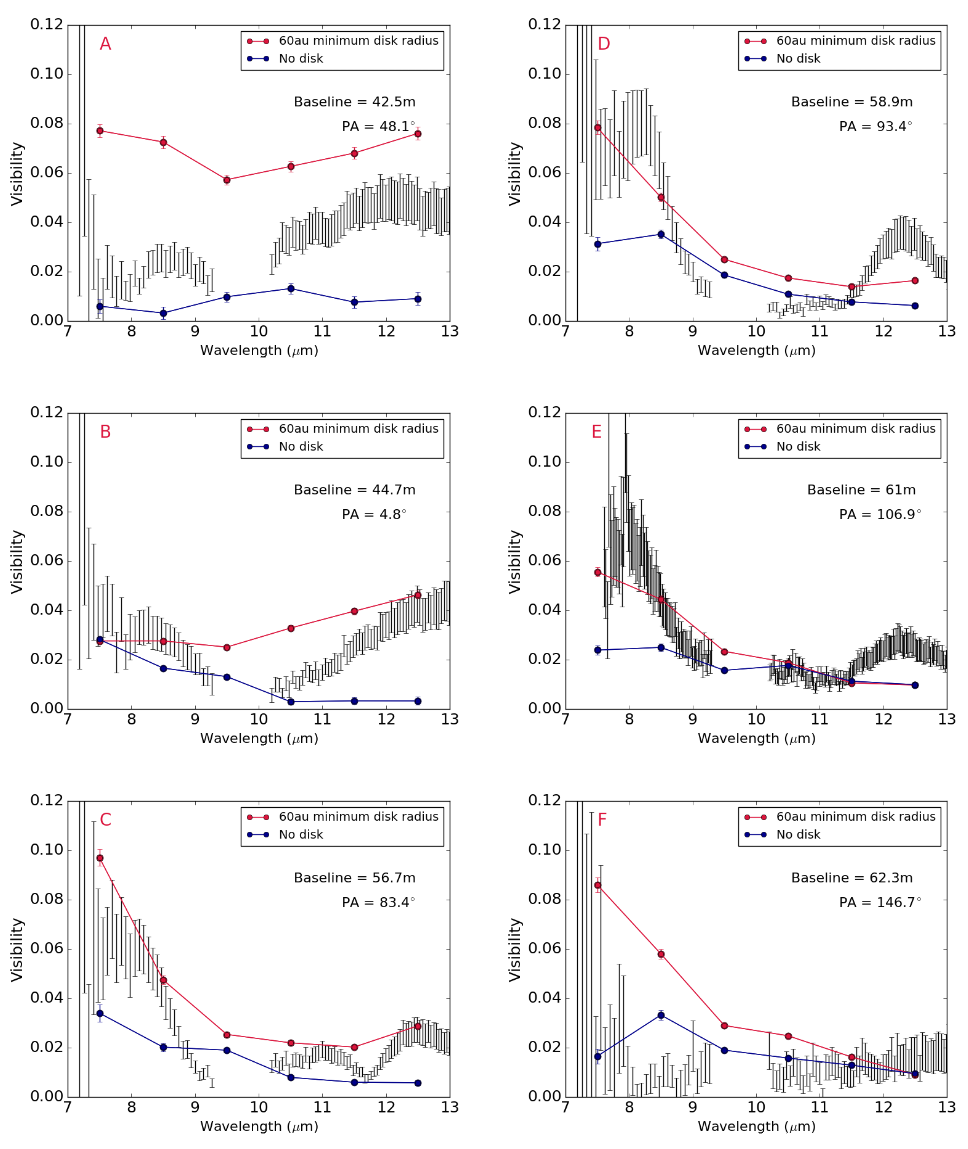}
\caption{Observed visibilities for each configuration (black) with the simulated visibilities for the best model fit (red) and a model with no-disk (blue).}
\label{nodisk}
\end{figure*}

\renewcommand{\thefootnote}{\fnsymbol{footnote}}
\footnotetext[4]{IDL is a trademark of Exelis Visual Information Solutions, Inc.}

\subsection{Testing the possibility of a bloated protostar}

While the central luminosity of the protostar is crucial to successfully fitting the SED, the radius and temperature of the central protostar can still be varied whilst maintaining this value. One object that creates the required luminosity is a central object with a temperature of 35000K and radius of 6R$_{\odot}$. Such properties are typical of an O7 zero-age-main-sequence object as defined by the work of \citet{strai}. This central object is similar to the object used in \citet{wit10} for W33A, but is dimmer by nearly 30\% due to its smaller radius (6R$_{\odot}$ as opposed to 8R$_{\odot}$) for the same stellar temperature. One could assume that because of the hot nature of the object that the protostar should therefore be able to ionise hydrogen in its surrounding material and create a H{\sc ii} region, contrary to the literature mentioned in Section 1. It is possible to include a protostar that does not have this high temperature, thereby maintaining its luminosity, by `bloating' the star; expanding its radius and lowering its temperature and this possibility has been modelled extensively (\citet{hosokawa} and \citet{omukai}). Bloated protostars have been utilised in the fitting of other MYSOs, namely AFGL 2136 \citep{wit11} and M8E-IR \citep{linz09}. 

\citet{hosokawa} propose the bloating of central protostars as a by product of massive protostellar accretion where the protostellar radius swells to 30-100R$_{\odot}$. In order for this to occur, \citet{hosokawa} required envelope infall rates of order 10$^{-3}$M$_{\odot}$yr$^{-1}$, which are comparable to our model envelope infall rate of 7.4$\times$10$^{-4}$M$_{\odot}$yr$^{-1}$. As such the bloating effect Hosokawa postulates is feasible according to our analysis, if the protostellar system manages to retain a large portion of its envelope infall rate and some accretion continues across the 60au gap.

Two models were run where the compact protostar was replaced with two different bloated protostars to assess their affect on our observables. Each bloated central object had the same luminosity as their hotter compact counterpart with the first having a radius of 30R$_{\odot}$ and a temperature of $\sim$16000K and the second a radius of 100$R_{\odot}$ and a temperature of $\sim$9000K, in agreement with the radii estimates of \citet{hosokawa}. Changing the radius of the central protostar while maintaining its luminosity has negligible effects on the VISIR profile but the silicate absorption feature of the SED did becoming marginally deeper for the 100$R_{\odot}$ bloated star. In terms of the MIDI visibilities, differences were seen but the visibilities did not go up and down uniformly, with increases and decreases seen across the same N-band spectrum for each configuration. The 100$R_{\odot}$ protostar showed an average difference in visibilities of 0.0025 compared to the compact object, meaning a percentage difference of $\sim$6\%. This did not produce significant changes except in the case of Configuration B, where the visibilities at 9.5 and 10.5$\mu$m decreased, improving the fit. For a 30$R_{\odot}$ bloated star the average change in visibilities was 0.0018, constituting a $\sim$4\% change in visibilities compared to the compact central object and the fit of Configuration B was again a slight improvement on the fit for the compact source, although to a lesser degree than the 100$R_{\odot}$ protostar.

\subsection{The weak silicate absorption feature in the SED}

The weak silicate absorption and the adjustments required to fit it present further questions about this MYSO. MYSOs are typically extremely dusty environments (with previous works finding deep silicate absorption features in other MYSO environments) yet the feature of G305 is weak. This implies a less dense surrounding environment and indeed, lower cavity densities were required in order to fit the silicate absorption feature of our observational SED. A weak silicate absorption feature can be a result of a combination of emission and absorption in the environment \citep{whittet}. \citet{levenson} for example note that a deep silicate absorption feature should be the result of a continuous optically and geometrically thick dusty medium while a clumpy medium whose curves are illuminated by from an outside source will have a much weaker feature as the emission compensates for the absorption. We cannot replicate such a clumpy medium in our modelling but it is possible that the real-life geometry of G305 could be suited to this case. We have mentioned that there is a source that is bright at longer wavelengths which could be illuminating the object and contaminating our flux measurements at longer wavelengths however this source is not bright in the mid-infrared and so one would not expect this illumination to be filling in the absorption feature. This combined with the fact that a low density environment is required for successful VISIR fitting of G305 contradicts the first case suggested by \citet{levenson} and implies that the weak silicate absorption feature is indicative of the nature of the source, not that of outside illumination, and that this MYSO is less dense than typical examples of its class. YSOs are believed to disperse over time, so the low density environment of G305 implies that this dispersion has begun, and that the MYSO is evolving from an envelope-dominated to disk-dominated system. This is supported by work by \citet{walshy}, who conclude that the lack of HC$_3$N, NH$_3$, OCS, or H$_2$O emission observed by ATCA implies that G305 has evolved to the point where it has cleared a significant portion of its surrounding material.

\subsection{The inner cleared regions}

The disk of G305 has an inner rim of 60au as opposed to the $\sim$6-11au inner radius of the accretion disk found by \citet{kraus} - almost an order of magnitude difference in size. In order to form a minimum dust radius of 60au through dust sublimation alone, one requires a central object of $\sim$5$\times$10$^{5}$L$_{\odot}$. When such a central source is included in the code the SED fit is violated. The sublimation radius of the central YSO that best fits the SED is $\sim$18au. However, a model with a sublimation radius of this size results in a much worse fit for the MIDI visibilities (with no change to the VISIR and SED). This implies that the inner disk has been cleared out to the 60au inner radius which best fits the observations. 

Such a clearing in the inner regions of the disk has been observed in lower mass protostars (see \citet{cieza}, \citet{vandm} for examples) and such disks are referred to as `transition disks'. \citet{wyatt} describes the first stage of the transition disk as a phase where accretion is ongoing, with the presence of gas but absence of small dust. This is in agreement with the dust types of the successful model which do not contain small grains or PAHs. \citet{tang}, \citet{vdm}, \citet{vdp} and \citet{fed} show inner radii of order 10s of au for low-mass transitional objects, in agreement with the size of the inner radius of our best-fitting model. Considering these properties and the weak silicate absorption feature discussed previously, our work suggests that the disk surrounding G305 could present an example of a transition disk around an MYSO. Such a disk has not been reported around a high-mass object and if the transitional nature of this disk can be confirmed this has implications for the massive star formation process, in particular its similarity to the evolutionary sequence of low-mass objects, presenting a new observed evolutionary phase for massive forming stars.

An additional disk phase is also proposed for lower-mass stars known as the pre-transitional disk. During this phase a very small gaseous inner disk (scales of order 0.1au) remains close to the inner star \citep{esp10} as material is removed at larger radii, so the inner cleared region is referred to as a gap. \citet{jilee13} find their CO band-head emission data of G305 is satisfied by a disk of 0.6$\pm$0.3au minimum radius, suiting this specification. Further observations of the gas at small radii could confirm whether the disk found in \citet{jilee13} and our work coexist and whether G305 presents an example of this evolutionary stage.

We cannot confirm the exact mechanism that has caused the 60au radius gap in this disk but various mechanisms for the clearing of the inner region have been proposed for low-mass stars. One such mechanism which translates well to the massive protostellar case is photoevaporation. Photoevaporation is a process whereby high-energy radiation fields heat the uppermost layers of the disk, raising the thermal velocities of its particles above the escape velocity \citep{hollenbachgurrrl} resulting in a disk wind. Given that MYSOs have typically large luminosities it seems reasonable to consider that this may be the source of clearing for G305. However it is the gas disk that exists within the dust sublimation radius that will be expelled by photoevaporation and this is not considered by a radiative transfer model, although the work previously mentioned by \citet{jilee13} implies that a gas disk still exists close to the star. Disk theory work suggests that material captured within the resultant disk wind can either be removed from the system or be recaptured and rejoin the disk at larger radii. It has been proposed that the mass-accretion rate decreases with age and will eventually become equal to the amount of mass loss from the disk \citep{clarke}. After this point, matter is solely lost through the disk wind and is not replenished, meaning that as material from the inner regions continues to accrete the inner regions begin to clear of material. As photoevaporation continues the inner hole can continue to grow as the rim is irradiated \citep{armie}. This model predicts low mass-accretion rates and disk masses and is therefore considered an unlikely mechanism for the formation of low-mass transition disks, however this has not been considered for MYSOs where the relative mass of the disk compared to the star is much smaller than in low-mass cases (for our model of G305 the disk is 5\% of the mass of the central MYSO).

\renewcommand{\thefootnote}{\fnsymbol{footnote}}
\footnotetext[5]{\url{https://exoplanetarchive.ipac.caltech.edu/}}

Another potential source of clearing proposed for low-mass disks is viscous evolution. The theory postulates that as gas falls towards the central protostar the remainder of the disk spreads outwards to larger radii to conserve angular momentum \citep{esptrans}. This gas is thought to be `driven' and in the low-mass case this is attributed to magnetohydrodynamic turbulence, which is not necessarily expected around MYSOs who display little magnetic activity. \citet{arce} however do note the presence of turbulence within and resulting from massive protostellar outflows at varying scales and as such we note that this mechanism may also present an explanation for the cleared inner regions.

Another mechanism discussed in the context of the cause of the hole in transition disks around low-mass stars is the presence of planets. The detections of planets around higher mass stars are few and tenuous. The NASA Exoplanet archive\textsuperscript{\P} provides a comprehensive list of companions detected around other stars. 3826 companions are listed in total and of those objects 3053 have listed stellar masses. From that subset, only 7 objects (0.229\%) are 7 solar masses or larger and the error bars on the determination of these stellar mass are vast and they could, in fact, be low-mass stars. Given that massive forming stars are thought to reach the main sequence in $\sim$10${^5}$yrs and that planets were not thought to form around low-mass until the late stages of their formation (between $\sim$10$^{6}$-10$^{7}$yrs) it would seem that planets could not form around massive stars. However, more studies are appearing that observe gaps and rings in low-mass protoplanetary disks at the Class 0/I stages. For example, \citet{sheehan} find that the low-mass Class I protostar GY 91 has multiple rings within its disk and conclude that if planets were sweeping out these rings that they must be able to form within 0.5Myrs of the disk appearing. \citet{harsono} find this implied through the absence of carbon monoxide isotopologues. They conclude that shielding by millimetre-size grains is responsible for the lack of emission and suggests that grain growth and millimetre-size dust grains can be spatially and temporally coincident with a mass reservoir sufficient for giant planet formation. \citet{manara} note that the protoplanetary disks observed around young stars do not have enough mass to form the observed exoplanet population and propose that one implied solution to this problem is that the cores of planets must form in less than 0.1-1Myrs. If the implications of these works are correct, then planets may be able to at least begin to form around massive forming stars and therefore disrupt the disk at small radii. It is unlikely that these protoplanets would survive once the central YSO reached the main sequence, therefore explaining the lack of confirmation of exoplanets around massive stars. 

A potential scenario that perhaps translates better to the massive star formation scenario is the possibility that a forming companion star could be sweeping out the inner regions of the disk instead of a planet. The multiplicity fraction for massive forming stars is estimated to be large, although the specific formation mechanisms of these binaries/multiple systems is not confirmed. \citet{sausage} include G305 in their pilot survey of binaries and do not class it as a multiple, but their method is suited to detecting wide binaries (separations of 400-46000au) and as such a companion star could be forming within the determined 60au radius which they do not detect. \citet{jilee18} detected a body orbiting around MM1 within the G11.92–0.61 high-mass clump using ALMA observations. They conclude that this object, MM1 b, could be one of the first observed companions forming via disk fragmentation based on the extreme mass ratio and orbital properties of the system. \citet{meyer} present 3D gravitation-radiation-hydrodynamic numerical simulations of massive pre-stellar cores. They find that accretion discs of young massive stars violently fragment without preventing accretion onto the protostars and that the migration of some of the disc fragments migrate on to the central massive protostar with dynamical properties comparable to a close massive protobinary system. Given that the mass of the potential disk around G305 is not well constrained by our infrared observations and that the disks observed around other massive stars (e.g. \citet{john}) have the potential to be very massive, a forming binary companion could therefore constitute a more realistic body to disrupt the disk at small radii. \citet{meyer} postulate that FU-Orionis-type bursts could happen at the same time as close massive binary formation, with \citet{car18} detecting such a burst around the MYSO S255 NIRS 3. The secondary source, G305B2, mentioned in Section 5.1, appears point-like in the single-dish images of \citet{mengyao} and could potentially be a forming `proto-companion'. If disk fragmentation is occurring, a piece of dusty material expelled from a fragmenting disk could also cause the presence of G305B2 and explain the poor fitting of Configuration A. Obtaining ALMA data for G305 could assist in confirming whether similar signatures to those observed around MM1a/b exist for G305 and therefore whether companion formation and disk fragmentation could be occurring.

Another possibility is that the winds arising from the MYSO itself could clear the dust within the inner 60au. While these winds would be a result of the luminosity and temperature of the object, which are input parameters within the Hochunk code, such winds are not a component of the models we use. \citet{parky} present hydro-dynamical models investigating how disk and stellar winds interact with envelope material and that reverse-shocks can occur at scales less than 500au. The work does not look at the 20-100au scales relevant to this work and given the limits of our model we cannot explain why the outflows expected to disrupt material at 500au would only disrupt material out to 60au as is the case for G305.

Given the uncertainty as to the similarity of low-mass and massive star formation we cannot conclude whether one of the discussed mechanisms or a combination of the effects is a more likely culprit for the cleared the dust within the inner disk regions of G305. Further observations and study will allow for a more detailed discussion of this subject.

\section{Conclusions}

Through the combination of multiple baselines of MIDI interferometric data, VISIR imaging data and an SED we constrain the characteristics of the massive YSO G305.20+0.21 with a 2.5D radiative transfer model. The best-fitting model indicates that the MYSO is surrounded by a large extended dusty envelope with bipolar cavities and a dusty disk, as per the Class I specification of protostars. Throughout the fitting process the following observations were made in regards to the diagnostic abilities of each type of observation:

\begin{itemize}
    \item The N-band visibilities trace the warm material within the inner regions of the disk and emission from the cavity walls.
    \item The emission at 20$\mu$m traces mostly cavity emission and is negligibly affected by changing disk parameters (for a distance of 4kpc).
    \item In order to successfully reproduce the observed weak silicate absorption feature in the SED a low-density protostellar environment is required.
\end{itemize}

This work joins the growing number of publications that suggest that massive stars do indeed form disks as a means of accreting material onto the central protostar. However, in order to successfully fit the MIDI visibilities the inner radius of the disk cannot be the sublimation radius (as would be expected) and a larger inner radius is required. This cleared inner region is archetypal of the more evolved `transition disks' studied around low-mass stars. Given the fact that successful fitting also requires a lower density, low extinction environment that implies an older MYSO, we propose that this disk could therefore present one of the first examples of a transition disk around an MYSO. 

\section{Acknowledgements}

We thank the anonymous referee for an exceedingly helpful report that allowed the improvement of this work. We wish to thank Katharine Johnston for her discussion in regards to disks around massive protostars and the radiative transfer modelling, Tom Robitaille for his discussions on radiative transfer modelling and Catherine Walsh, James Miley and Alice Booth for their discussion of disks in general. We also thank the STFC for funding this PhD project and ESO's Director General Discretionary Fund that enabled invaluable collaborative work. The modelling presented in this work was undertaken using ARC3, part of the High Performance Computing facilities at the University of Leeds, UK.

\bibliographystyle{aa}

\end{document}